\documentclass[manuscript]{aastex} 

\def\edcomment#1{\iffalse\marginpar{\raggedright\sl#1\/}\else\relax\fi}
\reversemarginpar

\begin{document}
\title{The Princeton Variability Survey} \author{Cullen Blake}
 \affil{Princeton University Department of Astrophysical Sciences,
 Princeton, NJ 08544 cblake@astro.princeton.edu }

\begin{abstract}
The Princeton Variability Survey (PVS) is a robotic survey which makes
use of readily available, ``off-the-shelf'' type hardware products, in
conjunction with a powerful set of commercial software products, in
order to monitor and discover variable objects in the night sky.  The
main goal of the PVS has been to devise an automated telescope and
data reduction system, requiring only moderate technical and financial
resources to assemble, which may be easily replicated by the dedicated
amateur, a student group, or a professional and used to study and
discover a variety of variable objects, such as stars.  This paper
describes the hardware and software components of the PVS device as
well as observational results from the initial season of the PVS,
including the discovery of a new bright variable star.
\end{abstract}

\section{Introduction}
	Amateur astronomers have long been involved in the study of
variable stars.  Organizations like the AAVSO (http://www.aavso.org)
and the Center for Backyard Astrophysics
(http://www.astro.bio2.edu/cba) have promoted the study of a variety
of phenomena by amateurs and overseen the collection of many years of
high quality data.  As technology, particularly the CCD, has become
accessible to a wider ranger of enthusiasts, the opportunities for the
amateur astronomer to conduct or lead scientific inquiry have grown
tremendously.  Recent projects such as TASS
(http://stupendous.rit.edu/tass/tass.shtml) and  HCRT
(http://www.mtco.com/\~~jgunn) have begun to make significant
contributions along these lines. The importance of a comprehensive
knowledge of variability across the entire sky has been outlined by
Paczy\'nski (2001) and explored experimentally through projects such as
ASAS (Pojmanski 1997, 1998).  Approximately $90\%$ of all bright,
existing variable stars are thought to be yet undiscovered, with ASAS
finding that approximately $3\%$ of a set of 140,000 monitored stars
turned out to be newly discovered variables with a wide range of
periodicities, as discussed by Eyer \& Blake (2001).  The potential
contribution from the study of a set of several thousand stars, even
for just a few nights, is enormous.  The ultimate goal of the PVS has
been to design a system to be utilized for just this purpose.  In
order for the desired scientific results to be achieved in as simple a
manner as possible, a number of criteria must be met.  Availability of
components, ease of implementation, and reproducibility of results
were all key concerns during the development of the PVS system.

Today, a wide array of advanced devices are available to the savvy
amateur and when these devices are combined with powerful and inexpensive computational
resources, the ability to survey the sky from a backyard is shown to
be well within reach.  The optical and electronic components of the PVS
are easily available through mail order, or even a local retailer, and
products already widely owned by amateurs were integrated whenever
possible. Though the commercial hardware systems utilized in the PVS
are shown to perform well, the power of the PVS design is derived from
the software that controls all aspects of data collection and actively
compensates for mechanical and electronic shortcomings in the hardware
components.  This collection of commercially available software
products, all adhering to a set of inter-program cooperation
guidelines called ASCOM (http://ascom-standards.org), work together on
a PC, under Microsoft Windows, to control a telescope, camera, and, if
required, a telescope dome.  These programs allow simple procedures
to be controlled from a Graphical User Interface (GUI) and for
complex observing sessions and data reductions to be scripted in Perl,
Visual Basic Script, or C++ languages.  This paper provides
descriptions of the hardware and software utilized in the PVS so
that a similar system may be constructed by interested parties.
Examples are presented of the type and quality of results that may be
expected.

\section{Hardware}
\subsection{Telescope}
	The PVS optical system centers around the popular Meade LX200
Schmidt-Cassegrain telescope (http://www.meade.com).  This telescope
is well known for its low price and ease of use,
 and was found to be
of more than adequate optical and mechanical quality for the purposes
of the PVS.  Here, the 12" aperture model was utilized, with the
standard f/10 to f/6.3 focal reducer, yielding a working plate scale
of 82"/mm.  In practice, any of the Meade LX200 Schmidt-Cassegrain
telescopes could be used. The PVS LX200 is permanently mounted on a
pier in equatorial configuration. Care was taken to ensure that the
telescope base was level and properly aligned with north celestial
pole, following the instructions in the Meade manual.  Although the
instrument can be operated in an ALT/AZ configuration, field rotation
would be problematic.  The telescope comes standard with the ability
to communicate with a nearby PC via an RS-232 to serial port
connector, facilitating the development of a robotic system.  This
communications cable can be easily assembled following instructions in
the appendix of the Meade manual and requires a four-wire RJ11
standard phone cable and an RJ11 to DB9 (9 pin serial) adapter.  The
voltages associated with this RS232 signal are very low, so it was
found necessary to utilize a high quality shielded twisted pair phone
cable, available at computer stores as a high speed modem cable.  With
this connection, encoders on both axes can be read out to the computer
and dual axis motors actuated in response.  The LX200 makes use of a
large Right Ascension gear of 180 teeth with a period of approximately
8 minutes.  The periodic error of this gear was found to prohibit
unguided exposures of over 60 seconds in length, with peak to peak
deviations of approximately $20\arcsec$.  The telescope does have the
ability to actively compensate by "learning" these errors through a
training process during which an observer manually maintains the
position of a guide star while the telescope's small integrated computer
records these adjustments.  The periodic error of the LX200 was not
found to be reproducible between pointing positions or nights, and
random tracking errors of $\approx10\arcsec$ remained even after
training.  An autoguider was utilized to compensate for these tracking errors during exposures at a rate of approximately 1Hz,
resulting in tracking accurate to the pixel level.  The necessity of an
autoguider does impose a limit on the percentage of the entire sky
that is observable by PVS. Since a reasonably bright guide star is
required and in general a rotation of the camera is not possible
during observations, only fields a fixed distance away from bright
stars were observable.  It was found that stars bright enough for guiding were abundant enough so that $\approx$1/3 of all blind
pointings happened to produce a usable guide star.

When permanently mounted in equatorial mode, the LX200 requires only
one alignment star in order to determine position.  Once the
instrument is aligned to a bright star of known coordinates, the
encoders will keep track of the telescope's Right Ascension and
Declination with reasonable accuracy for the duration of the night's
observations.  Through the RS-232 connection, the controlling PC can
slew the telescope toward any set of coordinates. Pointings across the
entire sky were found to be accurate to within $8\arcmin$, placing the
desired object well within the field of view of the CCD camera used by
PVS.  Locally, the pointings were further enhanced to much higher
accuracy.  With the assistance of software, pointings to within a few
arcseconds were routinely achieved during the data taking process.  An
important benefit resulting from the popularity of the Meade LX200
model is the Meade Advanced Products Users Group
(http://www.mapug.com), an on-line reference where hundreds of
technically minded amateur astronomers have posted their experiences
and insights into LX200 usage, design, and enhancement.  One major
problem with the LX200 optical design is the focusing mechanism.  The
focus is controlled through movement of the primary mirror, resulting
in perceptible image shift and very poor focus repeatability.  A new
focuser, model NGF-2 manufactured by Jim's Mobile Instruments
(http://www.jimsmobile.com), was added to remedy this problem.  This
motorized focuser, with an analog focus micrometer, was found to be
very rigid and reliable.  This addition adds several centimeters to
the optical path length, resulting in slight vignetting at a level
$<10\%$.  Optical filters of the 50mm round variety were inserted
into the optical path inside the focuser, as opposed to closer to the
CCD chip, in order to help reduce vignetting effects and to ensure that
any blemishes on the filter surface were well out of focus.  Optical
filters manufactured by Omega Optical (http://www.omegafilters.com)
were utilized.  The system accommodates only a single filter at a
time, but the filters may be easily switched by removing the CCD
camera from the focuser. The focuser was attached to the rear cell of
the telescope using a special wide aperture adapter plate purchased
from Jim's Mobile Instruments. To the front of the focuser was attached
the focal reducer, and finally the camera was affixed with a T-thread
adapter to the focal reducer.  Care was taken to ensure that the camera
was aligned as closely as possible with the axes of the telescope in
order to simplify guiding by aligning the rows and columns of the CCD
chip with North, South, East, and West.

\subsection{CCD Camera}
	The advent of the low cost CCD camera has revolutionized
amateur astronomy.  Technology once available only at professional
observatories is today within the reach of many dedicated home observers.
Several companies manufacture such CCD systems, but SBIG
(http://www.sbig.com) was chosen as a result of experiences with
cameras from a variety of makers and as a result of SBIG's
reputation for innovation, quality, and good customer service. The
model ST-8E was selected as a reasonably priced, large format camera
accessible to amateurs and schools.  The 16-bit ST-8E uses a Kodak
KAF-1602E CCD chip with an area of 1530 by 1020 pixels, each
$9\micron^2$.  This chip reaches a peak quantum efficiency of
$\approx70\%$ at a wavelength of 575nm and slowly loses QE to
$\approx10\%$ at $1\micron$ wavelength. An effort was made to
calculate the throughput of the PVS system by observing the Vega-like
star HIP 25593 (AOV, $m_V=10.55$). A result was obtained 
comparable to an estimate based on the efficiencies reported by the
manufacturers of the telescope, CCD, and filter. Based on a calculated
extinction coefficient of 0.09 mag/airmass in the I band, the Vega
flux standard at 555.6nm of $3.44\cdot10^{-9}ergs/s/cm^{2}/\mathrm{\AA}$ from Hayes
(1985), and an assumed blackbody spectrum for HIP 25593, the total
system efficiency was calculated to be $9.95\%$. Using the published
efficiencies of the CCD, telescope, and optical filter, and assuming
the unknown efficiency of the focal reducer to be $\approx80\%$, a
total system efficiency of $10.7\%$ was estimated.  

Located adjacent
to the imaging CCD on the same focal plane, is a smaller TI TC-211
chip used for autoguiding. Both chips are cooled with a single stage
thermoelectric device to approximately $25\degr$C below ambient.  At a
chip temperature of $0\degr$C, the dark current is measured to be
$1e^-/pixel/s$.  The camera is mechanically shuttered, so the full
Kodak CCD is used for imaging at all times. The ST-8E comes with the
option of antiblooming circuitry which helps to lessen the effects of
bleeding caused by an overexposed stellar image.  Many amateur
astronomers choose this option since it is helpful in deep sky
imaging.  Unfortunately, the antiblooming reduces the full well
capacity of the CCD from $100,000e^-$ to $50,000e^-$ and may also be
responsible for non-linear effects that are seen to occur as a pixel
nears full well capacity.  For use with the PVS, a non-antiblooming
version of the camera was chosen.  The ST-8E reads out to the
controlling PC via a parallel port interface in approximately 52
seconds, with a read noise of $15e^-$RMS.  In conjunction with the
previously described PVS optical system, the ST-8E yields a field of
view of approximately $23\arcmin$ by $15.5\arcmin$ at a plate scale of
$0.91\arcsec/pixel$ with the use of the 0.63x focal reducer.  All
electronics are integrated into the head of the camera unit, so its
total dimension is approximately 12cm round with a total weight of
0.9Kg. This weight is positioned several centimeters from the rear of the
optical tube, and must be balanced with a counterweight placed near
the front of telescope.

\section{Astronomy Common Object Model software}
	The computer control systems for the PVS instrument were all
run and developed under Microsoft Windows 2000.  Windows was chosen
since it is far more accessible to the majority of amateur
astronomers.  The programs used by the PVS are all commercially
available products that adhere to the guidelines of the Astronomy
Common Object Model (ASCOM, Medkeff 2000).  ASCOM is a new and
extremely powerful software concept that utilizes Windows ActiveX
objects each representing an instrument component or a mathematical
manipulation, in order to integrate the functions of a variety of
independent software packages and facilitate the automation of an
observing system.  With products that incorporate ASCOM, complex
scripts may be written that may control an entire night of
observations with data handling and control of the telescope, CCD, and
dome.  The ASCOM products also have the advantages of open source,
since scripts with differing functions can be easily written and
traded between observers.  It was felt that the ASCOM initiative
represented a new direction in amateur astronomy, and much of the PVS
was designed around the usage of ASCOM compliant products. While ASCOM
products were used here to control a Meade telescope and SBIG camera,
the same programs and scripts may be used with a variety of different
CCD cameras and telescope mounts.  The telescope control software
ACPv2.0 (http://acp.dc3.com) was chosen to oversee telescope motion
and pointing as well as coordinate and time transformations.  The
astrometric software PinPointv3.0 (http://pinpoint.dc3.com) was
selected to manage all astrometric calculations.  A wide variety of
CCD camera control products are currently available, but
MaximDL/CCDv2.12 (http://www.cyanogen.com) was found to be generally
superior.  MaximDL/CCD was used to control the camera and manipulate
FITS data.  Since all three of these products are ASCOM compliant,
scripts may be prepared that utilize functions from each in order to
carry out observations or analysis.  A variety of languages may be
used to prepare such scripts, but Visual Basic Script was selected as
the PVS language since it is relatively easy to learn and is an
integrated feature of Microsoft Windows.  Scripts may be prepared as
plain text documents in Notepad and executed with Microsoft Internet
Explorer or from within the GUI of ACP.  The data taking procedure was
managed by a master script that incorporated elements from all three
software components.  ACP was used to send movement commands to the
telescope, calculate topocentric coordinates for objects based on USNO
calibrated UTC time and the telescope latitude, longitude, and
elevation. ACP also handles issues related to time, such as Julian date
calculations.  PinPoint created full World Coordinate System (WCS,
Calabretta \& Greisn 2000) astrometric solutions for CCD data images
based on best guess center-of-frame coordinates from ACP and a CDROM
version of the USNOSA2.0 star catalog.  These fits had typical
residuals of less than a pixel, and were utilized to update the
exact pointing of the telescope in near real time. Once the telescope
was positioned close to a desired set of coordinates by ACP, an image
was taken, the exact pointing was  calculated by PinPoint, and then the
telescope was moved to compensate for the error (up to a several
arcminutes) in the initial pointing.  This entire process of
point-image-solve-repoint takes approximately 20 seconds when a 3x3
binned image is used for the test pointing and the process is executed
on a computer with an 800Mhz Celeron CPU. On this same computer,
PinPoint is able to solve a full resolution 1530 by 1020 image in $\>15$
seconds.

The CCD camera itself was controlled through the program
MaximDL/CCD. This program sends commands to the camera and writes out standard 16-bit FITS files with
headers containing the usual information, as well as the WCS
astrometric structures prepared by PinPoint.  MaximDL/CCD also
controls the autoguiding by downloading images of a bright ($<10m_I$)
star placed on the secondary CCD chip, located adjacent to the main
chip, and sends small positioning corrections directly to the
telescope mount at rates of up to a few times per second.  These
corrections were adequate to alleviate most of the periodic error in
the Right Ascension gear of the telescope.  The coordinates of guide
stars were provided in the PVS scripts and with the known offset from
the center of the main CCD to the center of the guide CCD, ACP and
PinPoint were used to place the guide star in the center of the guide
CCD. At the beginning of the night the script incorporated a guider
calibration, whereby the exact alignment of the CCD pixels with the
directions of movement of the telescope, and the speeds of those
movements, are determined.  As long as the camera was not removed from the telescope, this
calibration was adequate for the night with small adjustments as a
function of Declination being automatically calculated by
MaximDL/CCD.  Lastly, attempts were made to use MaximDL, in
conjunction with ACP, to automatically estimate and control focus
values.  This was done by taking a series of 8 exposures of a bright
star and attempting to minimize the
stellar FWHM  while ACP slowly changes focus in steps of 0.005cm.  In practice, the large stellar FWHM and highly variable
seeing made reliable focus determination very difficult. While the PVS
system was being developed, a wide range of ambient temperatures were
encountered, but the air temperatures during the time when the
variable star data was taken remained fairly constant. The air
temperature outside of the dome typically fell by $6\degr$C during the
night, resulting in a focus change of $\approx-0.003cm/\degr C$. This
rate was calculated during very stable observing conditions. On a more
typical night, the focus values were found to be repeatable at the
+/-0.005cm level, meaning adjustments were only practical when
large changes in the outside temperature were encountered. After much 
experimentation, it was found that it was generally most efficient to focus 
once per night, at about 3 hours after sunset, using ACP and MaximDL to 
automatically estimate the best nightly focus value.

\section{PVS Implementation}
\subsection{Site Description}
	The PVS data was taken from a dome located on top of Peyton
Hall, Princeton University, Princeton, New Jersey.  This suburban
locale provided less than ideal observing conditions, but conditions
typical of those endured by many amateur astronomers.  Although the
sky was seldom, if ever, photometric, and seeing averaged approximately
$5\arcsec$, such conditions may still produce accurate differential
photometry.  The most significant limitations to the observations were
sky brightness and poor seeing.  Since the sky above
Princeton is light polluted, observations were taken in the
Bessel I band.  This bandpass is red enough to be free from much
of the light produced by nearby academic buildings and major
metropolitan centers as well as much of the light from the moon and
falls on an area of high CCD efficiency.

Sky brightness in I band is  related to atmospheric glow
rather than man made light pollution.  In I band, the sky brightness
above the PVS observing sites was found to average 17.7
mag/arcsecond$^2$.  Teare (2000) compiled sky brightness for a number
of sites.  For comparison, the measured brightness at Mt.Wilson,
California, located above the Los Angeles basin and its millions of
inhabitants, is reported to be 18.8mag/arcsecond$^2$ in the direction
of the city.  The sky brightness at CTIO, Chile, is reported at 19.9
mag/arcsecond$^2$ in I band.  The average I band sky brightness for
Princeton, Mt.Wilson, and CTIO are found to be more similar than the
average sky brightness in other bands, indicating that the I band sky
is less dependent on man made light pollution than other bands. The
Princeton site is brighter in I band than Mt.Wilson probably as a
result of the differences in altitude and atmospheric moisture between
the two sites.  Figure 1 presents a histogram of sky brightness at the
PVS site over a month of observations.  With exposure durations
reasonable for a survey ( up to a few minutes in length), the limiting
magnitude with the PVS 12'' telescope at Peyton Hall was found to be
approximately 16 in I band .  The atmospheric conditions in Princeton
were also observed and found to be notably poor. As a result of a
combination of dome seeing and atmospheric turbulence, compounded by
guiding errors, the image FWHM was found to routinely be greater than
$6\arcsec$. Figure 2 presents a histogram of image FWHM values during a 
month of PVS observations. Scintillation effects were severe, and
caused large amplitude (up to $30\%$) variations in image FWHM on
frequency scales up to the 5Hz.  At least a portion of the seeing
effects were attributable to heat rising up into and around the
telescope dome from Peyton Hall below, but similarly bad seeing
conditions are reported from other observatories in the area. As an
example, it was found that seeing was often $1\arcsec-2\arcsec$ worse
than average on the coldest nights, when the heating system of Peyton
Hall was activated. To this end, great care was taken to ensure that
the dome was adequately ventilated and temperature equalized with the
outside. While the use of large fans to helped to improve
the results obtained during the first hour of observations, the
average seeing values achieved during the middle of the observing
night were found to be unaffected by efforts to increase the
efficiency of temperature equalization in the dome.  The performance
of the PVS system as tested in Princeton is  limited by poor
conditions, and thus, in a better climate the capabilities of this
instrument might be considerably extended.

\subsection{Data Collection}
	With the objective of monitoring several hundred stars, an
area of the sky was chosen near the galactic equator, in the
constellation Auriga, which would be visible for many hours during the
period of the PVS observations.  Sections of sky near the Milky Way
should in theory yield the highest number of stars per image, but any
area of the sky could, in principle, be studied.  A strip of 5
pointings, amounting to a total field of view of $0.5\degr^2$, was
selected near the intersection of the galactic plane and the
ecliptic. One of these pointings contained the open star cluster NGC
1912. This rich cluster, with $\approx300$ members, has an HR diagram
from the observations of Subramaniam \& Sagar (1999), which indicates
that it may contain lower instability strip variable stars at
magnitudes accessible to the PVS instrument.  A nearby open cluster of
similar age and metallicity, NGC 2516, was studied by Zerbi et
al. (1998) and was found to have several short period intrinsic
variable stars.  The other four fields, all within a few tens of
arcminutes, were picked  to
be fields with many stars from digitized sky survey plates.  Eyer \& Blake (2001) analyzed the variable
star catalog produced by ASAS and found that the histogram of periods
below 50 days peaked at periods a bit longer than 1d.  As a result, it
was not critical that the PVS data be time sampled at rates of more than
two or three times per hour.  For example, 15 data points per night
over several nights would yield frequency sampling perfectly adequate
for accurate determinations of the periods of many intrinsic (Cepheid,
RR Lyrae, dwarf Cepheid,Gamma Doradus) and extrinsic (contact
binaries, eclipsing binaries) variable stars.

Based on the conditions in Princeton, each of the five fields observed
by PVS was allotted 240 seconds per observation.  This amounts to 150
seconds of integration, 50 seconds for image read out, and 40 seconds
for overhead associated with astrometric analysis and
pointing.  The 150 seconds for integration time was found to balance
 the faint limit imposed by sky brightness and the saturation of
bright ($m_I<8$) stars.  The telescope was parked at the end of the
evening at a prescribed Altitude and Azimuth so that the next night,
with accurate local time and telescope latitude and longitude, a
reasonable estimate could be made of initial telescope Right Ascension and Declination. The night's initial alignment of the telescope
with a bright star was therefore automated. After the power was
turned on, a script wasexecuted and the remainder of the night's
observations were completed robotically.  Images of the five fields were
taken in series for a prescribed number of hours. For example, with
five fields to be observed, the telescope system would automatically
cycle between the fields in the pattern A,B,C,D,E,A,B... until morning
twilight is reached.  This strategy resulted in up to 20 images of
each field per night. The images were not intentionally dithered. Instead it
was intended that each each exposure of a given field place the stars
near the same pixels. A master script initialized the camera,
began the camera cool down, executed the initial telescope alignment,
moved the telescope with high accuracy to pre-determined
coordinates, began guiding, and gathered data exposures. Unfortunately,
this script did not include dome automation for the results presented
here.  The cost of automating the pre-existing dome at Peyton Hall was
 prohibitive, so the dome was moved by hand every hour
or so.  There are several commercially available robotic dome systems
that utilize ASCOM compliant software, and as a result, dome control
may be integrated seamlessly into PVS scripts in the future.  Sets of
five dark frames were taken each night and used during data reduction
to subtract both the dark current and bias from the data images. Sets
of 10 flat field exposures were taken from the evening sky following
sunset and from an evenly illuminated flat screen inside the dome and
were used to compensate for pixel to pixel variations in system light
sensitivity. Exposure times for the flat fields were set so that the
CCD chip was filled to approximately half of full well
capacity. Experience with both sky and dome flat fields showed that
the increased flexibility of taking flats inside the dome, as opposed
to off of the twilight sky, made this method advantageous. While the sky flats were
taken, it was the dome flats which were later utilized for the image
reduction since they were found to be more repeatable and could always
be produced, even if weather limited observations to only a small
number of hours during the middle of the night. A test was done to
determine the quality of the dome flats, as compared to the sky flats,
by reducing and photometering a small image section time series with
both types of flats. The differences in the flat fields was found to
have a negligible effect on the resulting photometry. The contribution
from the flat fields to the final photometric errors was estimated to
be $<0.5\%$.

\subsection{Data Reduction}
	As a result of the limiting magnitude at the PVS test site,
the data images were not considered crowded.  Typically, average star
densities on the CCD frames were on the order of 3,000 pixels per
star. In this regime, a number of different photometry techniques are
applicable (Alard 2000, Alard \& Lupton 1998). Though simple aperture
photometry could be utilized, the technique of differential image
analysis was applied in order to gain experience with this method for
the time when a better observing site might produce CCD images
sufficiently dense to warrant this advanced treatment of the data.
This method has several other advantages, including high accuracy, no
need for identification of individual comparison stars to produce
differential magnitudes, and an ability to deal with the highly
variable PSFs found in the PVS data.  Image subtraction has been shown
by Wozniak (2000) to produce photon noise limited photometry and thus
could be expected to yield the best possible results for the PVS data.
Data reduction tasks such as flat fielding, dark subtraction, and
cosmic ray removal were carried out within the Interactive Data
Language (IDL) environment.

There are numerous publicly available IDL codes to manage FITS file
I/O (http://idlastro.gsfc.nasa.gov/homepage.html) as well as a wide
variety of mathematical and image analysis routines.  Each night's
flat fields were read into an image cube, individually normalized by
dividing by the sigma-clipped image mean, and then the normalized
images were averaged, with sigma clipping, to produce a robust nightly
flat field. While there were enough counts in the flat fields to limit
the statistical noise from the flat to much less than the statistical
noise from the sky background in the images, systematic effects were
visible in the flat fields as subtle structural changes between
images.  Fortunately, star centroids remained quite close to the same
pixels throughout each night, so large scale, high amplitude
gradients and inconsistencies in the flat fields were effectively
rendered null for differential photometry.  The dark frames were also
read into a data cube and averaged, in order to help eliminate cosmic
ray hits, and a nightly dark frame is created.  The dark current of
the CCD camera is a function of chip temperature, so it is important
to prepare a separate dark frame for each night's observations.
Similarly, changes in the optical surfaces and focus require that a
separate flat field be prepared for each night.  The dark frame was
subtracted from the image frame and the result divided by the flat
field.  Finally, cosmic ray hits were removed from the data images.
There are several available IDL codes to do this, but LACOSMIC
(van Dokkum 2001) was found to be most effective.  New FITS files with
with the reduced data and full header, including WCS information, were
written.  A file called $\it{dates}$ and containing the image names,
Julian date times of observations, and seeing values was also written
out to be later referenced during the photometry.  The seeing, or
image FWHM, was calculated by fitting a simple Gaussian to the
profiles of isolated stars.  The reduced image FITS files and the
$\it{dates}$ file were then used with the image subtraction code
to produce photometric data.

	The image subtraction algorithms were implemented with the
ISIS2.1 package prepared by Christophe Alard
(http://www.iap.fr/users/alard/package.html).  The first step in the
analysis was the sub-pixel image alignment, which is required prior to
image convolution.  Since the CCD camera was occasionally removed and
re-attached to the telescope, it was necessary to use the 2nd degree
image interpolation option so that translations, as well as small
rotations, could be corrected for.  The image to which all others were
aligned, called REFERENCE in the ISIS configuration file, was chosen
to be the image with the best seeing.  It was important that this
fiducial image be of high quality so that the alignment would be
sufficiently accurate. ISIS created a file called $\it{loginterp}$
which tabulates the average residuals for each alignment with which the
sub-pixel residuals were verified.  Any image that failed to align to
the reference image would be flagged in the log file and would
therefore be removed from list of data images to be analyzed.  Once the images were well aligned, a
new reference image of high signal to noise was created.  To do this
the 10 best images were combined using the $\it{Simpleref}$ procedure.
The 2nd degree spatial kernel variation option along with 1st degree
background variation fitting were used in order to help compensate
for the visually obvious variations in PSF and background across the
frames. The images were not sub-divided prior to analysis as was
done by some users of the image subtraction method.  The reference
image was convolved to match the PSF within each data image and
then the two were subtracted to determine differences in flux.  The
subtraction was carried out by the $\it{Subtract}$ script and produced
subtracted images that were nearly devoid of stars.  The entire
process is computationally intensive and may take several hours,
depending on the number of data images.  Using the ISIS package on a 
Linux PC, the production of fluxes from approximately 100 images
required just less than 1 hour of computational time on a 1.0GHz
Pentium III work station with 1GB of RAM.  The errors in the
subtraction photometry were found to be small and well-behaved. Strictly speaking, variance in the subtracted images should
be the sum of the Poisson deviations in the reference and data
images. In practice, it was found that bright but not saturated stars
often have convolution residuals several times those expected from
photon noise.  This effect, due to seeing variations, is explained by
Alard and Lupton (1998).  The image subtraction methods were found to
work extremely well for the PVS data. Figure 3 shows the histogram of
the residuals of a subtracted image normalized by the Poisson errors
expected from the reference and data image. This figure demonstrates
that there was very little residual starlight in the subtracted
images.  The idealized distribution of the residuals, N[0,1], is
overplotted with the dashed line.  The actual residuals fall into a
N[0.005,0.98] distribution, indicating that the image subtraction
techniques worked quite well and that accuracies at the limit of
the sky noise should be achieved.

In order to detect variability, the subtracted images, normalized by
the gain adjusted sums of the Poisson deviations of the reference and
data images, were squared and coadded. The resulting variance image
made variable objects easily identifiable both visually and by a simple
detection script.  Variable sources, which have variance well above
that expected from just photon noise, were photometered with basic
aperture photometry on the individual subtracted images in order to
determine how stellar flux changed with time.  False detections were
found to occur for bright and nearly saturated stars as well as for
stars near the edges of the frames where drift in the image centroid
during a night may cause stars to come in and out of the field of
view.  The errant bright star detections, in this case for stars
brighter than about 8th magnitude, were generally disregarded.  In
practice, this amounted to very few stars, and considering that
many of the stars this bright would have been previously studied by
other programs, the loss is considered negligible.  The photometric
errors were found to be quite small for all stars. Figure 4 shows
average errors in magnitudes for a star of a given I band magnitude.
Overplotted on these points is the line indicating the magnitude
errors from sky noise that would be expected in the aperture
photometry of a subtracted image with average seeing and sky
brightness.  For stars fainter than approximately $m_I=11$, the
photometric accuracy was nearly sky limited. The contribution from
scintillation noise was calculated according to the formula of Young
(1974) and was found to be $\approx0.002$mag for 150s
integrations and the low elevation of the PVS site. 

A master list of stars for each frame was prepared from the averaged
reference image using the DAOPHOT FIND routine with detection
thresholds determined by Poisson statistics, as outlined in appendix B
of the DAOPHOT II manual.  The fluxes, as measured from the subtracted
images of all of the stars in the master list, were searched for
periodic signals using the Lomb Periodogram algorithm described by Lomb
(1976).  This algorithm produces a probability statistic that the
fluctuations of a given star are due to noise. A list of the 10 most likely variables in each field was prepared so
that these stars could be inspected visually. Fluxes were converted to
magnitudes by attempting to find stars of known I band magnitude in each
frame in order to set a frame zero point. The comparison stars for each
frame were chosen by searching the SIMBAD database
(http://simbad.u-strasbg.fr/sim-fid.pl) for stars with
known spectral type and magnitude. A total of 5 stars with
$m_{I}<11$ and spectral type A to F were found within the observed fields and used to set the magnitude
zero points. Since here the absolute zero points of the stellar
magnitudes are less important than the differential magnitudes and
cataloged stars with accurate I band magnitudes were difficult to
find, only these simple steps were taken to tie the PVS magnitudes to
the standard magnitude scale.  This process does introduce
errors in the zero point magnitudes of all the stars, but it in no way
affects the differential photometry. Since change in flux is being studied here, rather than average flux,  these errors are considered acceptable.

\section{Results}

Observations totaling about 30 hours were carried out during the month
of 2001 December.  Data was taken under a variety of conditions,
including partial cloud, since differential magnitudes are known to be
relatively robust to poor observational conditions. Even during the
best hours of atmospheric transparency and stability, the conditions
were found to be relatively poor. Seeing
and sky brightness combined to produce a limiting magnitude, at which
expected photometric errors due to sky noise just exceed $10\%$, of
approximately 15.5 in I-band.  The poor conditions also caused
occasional high amplitude scintillation affects that, when combined
with residual mechanical guiding errors, caused jumps in the guide
star centroid that were unrecoverable by the autoguider. As a result,
approximately 10\% of data images were smeared and not useful. These
images were rejected by the astrometric fitting routines. 
 The total number of stars observed was approximately 3,000
over the five fields, with approximately 1,000 of these stars being
bright enough to have S/N such that photometric errors are
$<10\%$. With a baseline of 30 days and data sampled at a incidence of 3
measurements per hour over portions of 10 separate nights, many types
of variable stars would be detected when enough stars are measured with
sufficient accuracy.  Experience with ASAS has shown that the rate of
short period eclipsing binaries is high enough that several may have
been expected in the PVS fields.  As a result of sky brightness and
seeing, the data images are unfortunately not deep enough or of high
enough signal to noise to produce these results.  Further more, many
of the variables discovered by ASAS ($3\%$ of observed stars)
vary on time scales longer than these observations, so they were not
detected with PVS. As a result of these factors, only one new variable
star could be positively identified at the several sigma level.  This
new variable, previously designated $BD+35\_1114$, is a 9.5 magnitude
O or B type star in the constellation Auriga, with J2000 coordinates
$05\fh28\fm09.6\fs +35\fdg16\farcm56\farcs$  The period was found to
be 0.81d and the amplitude 0.34 mag.  The phased light curve is shown
in figure 5.

\section{Conclusions}

The design of a robotic photometric telescope, built out of
commercially available and inexpensive components, is outlined.
Results from one month of observations with the PVS instrument are
presented, and it is demonstrated that the instrument produces high
quality photometric results, even from a location with extremely bad
atmospheric conditions.  The discovery of a new short period variable
star is presented.  Though the system performed as well as could be
expected given the atmospheric conditions at the Princeton, New
Jersey, site, the full potential of the PVS design cannot be
determined until it is tested at a site with better conditions.  A
decrease in sky brightness by 1 magnitude, an improvement of mean
seeing to 3'', and a higher percentage of clear nights would result in
a tremendous increase in the number of variable stars that could be
discovered by the PVS system.  Many suitable sites exists, including ones,
such as Mt.Wilson, within easy reach of major metropolitan areas of
the American South West. Instrumental and computational techniques
are developed and tested here so that a system  could be
efficiently and economically installed and operated at a better site.

\acknowledgements The author would like to thank Bohdan Paczy\'nski
for his tireless support, guidance, encouragement, and patience while
this research was carried out, William Golden for his generous support
of this work, and Bruce Draine for many hours of assistance and
discussions. Thanks also to Bob Denny, the developer of the ASCOM
standard. It is a pleasure to thank the referee, Michael 
Richmond, for his extremely detailed and helpful comments on this work and 
for his assistance in making important improvements to the original 
manuscript.

\begin{figure}
\plotone{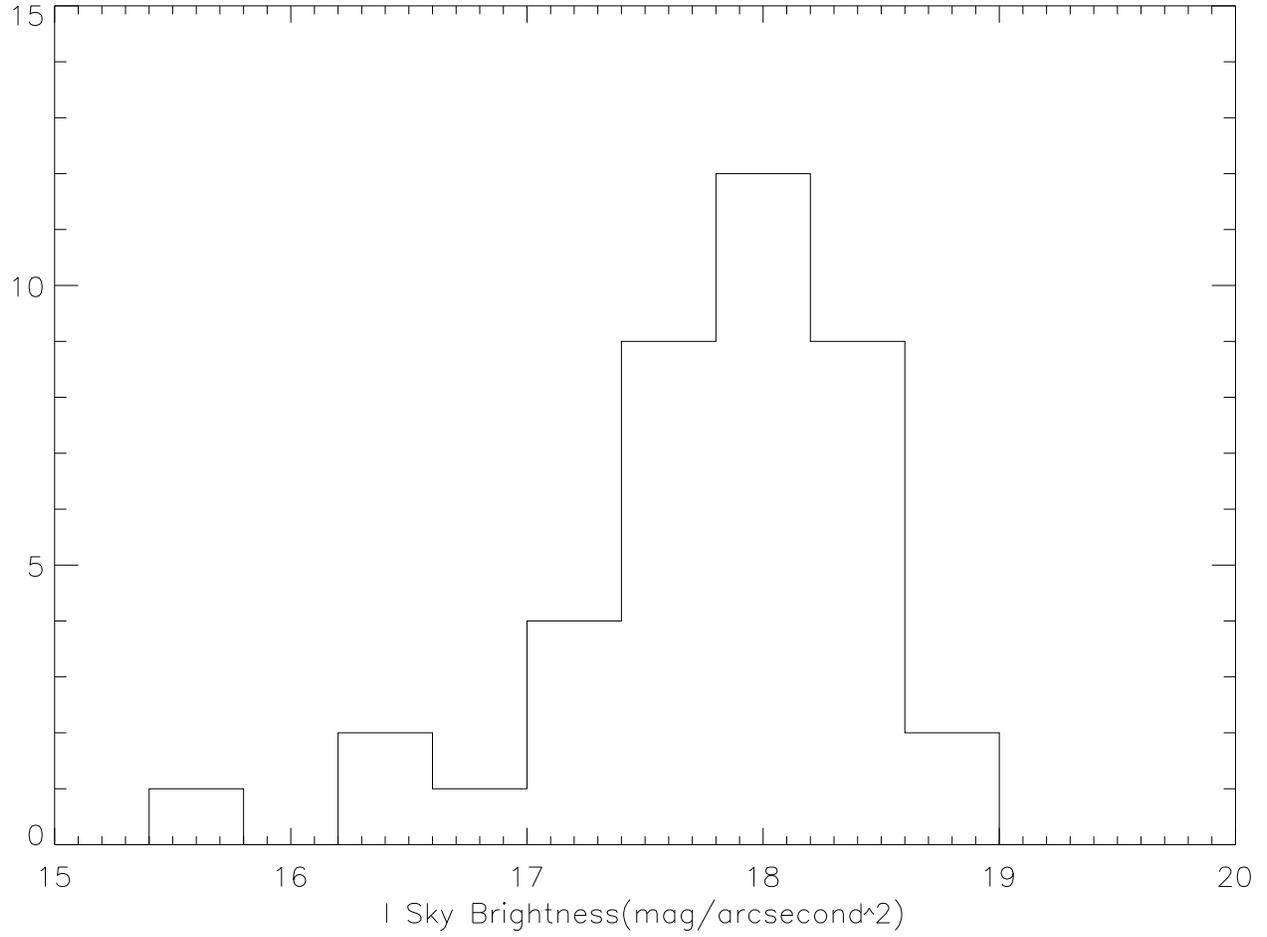}
\caption{A histogram of the I band sky brightness from the PVS site in
Princeton, NJ for the month of December, 2001. For comparison, the
mean sky brightness, in mag/arcsecond$^2$, at CTIO is 19.9}
\end{figure}

\begin{figure}
\plotone{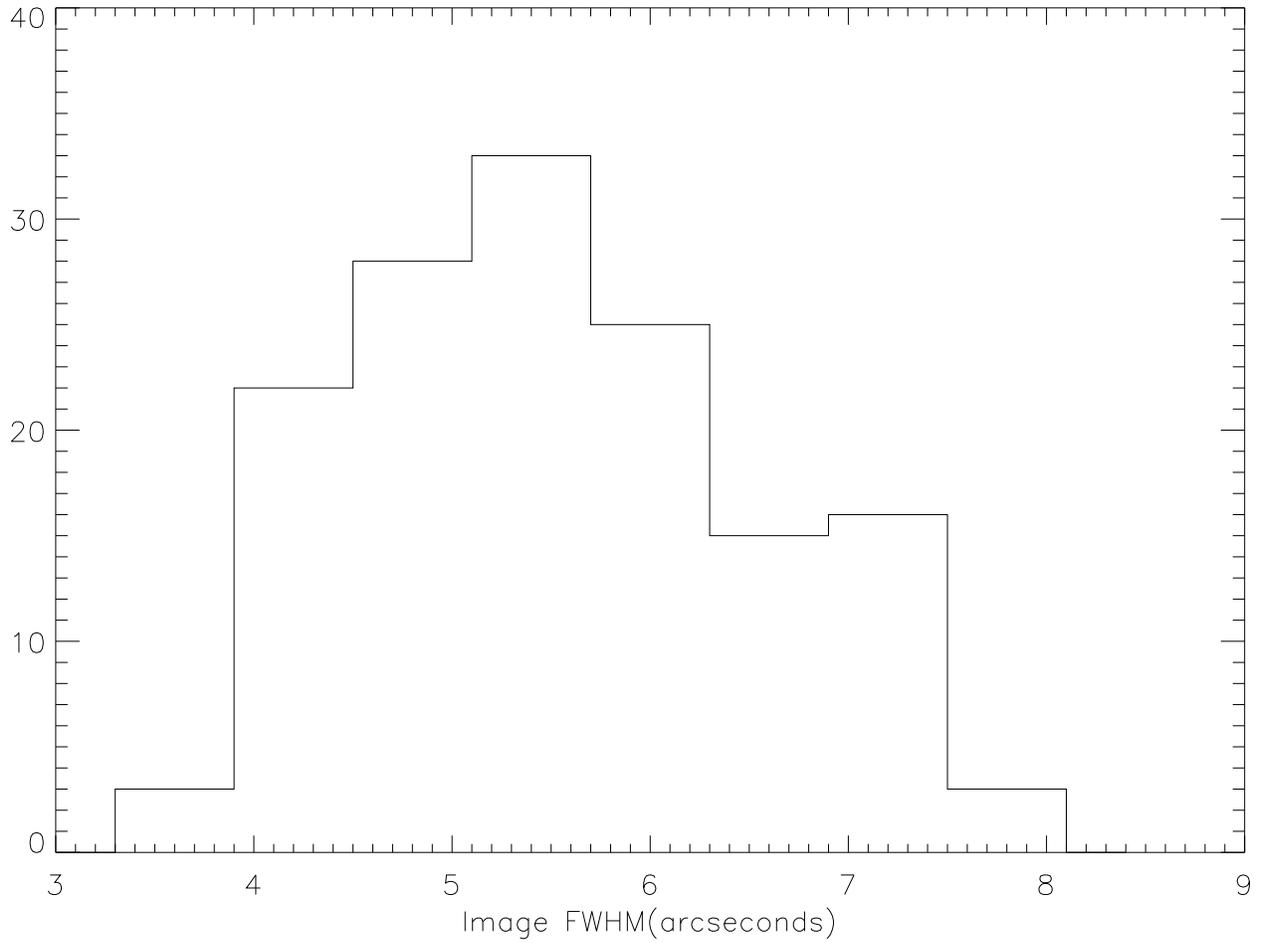}
\caption{A histogram of image FWHM at the PVS site in Princeton, NJ,
for the month of December, 2001.  Seeing is clearly poor, but the
the  atmospheric seeing is almost certainly
compounded by dome seeing and guiding errors to produce these
large PSFs.}
\end{figure}

\begin{figure}
\plotone{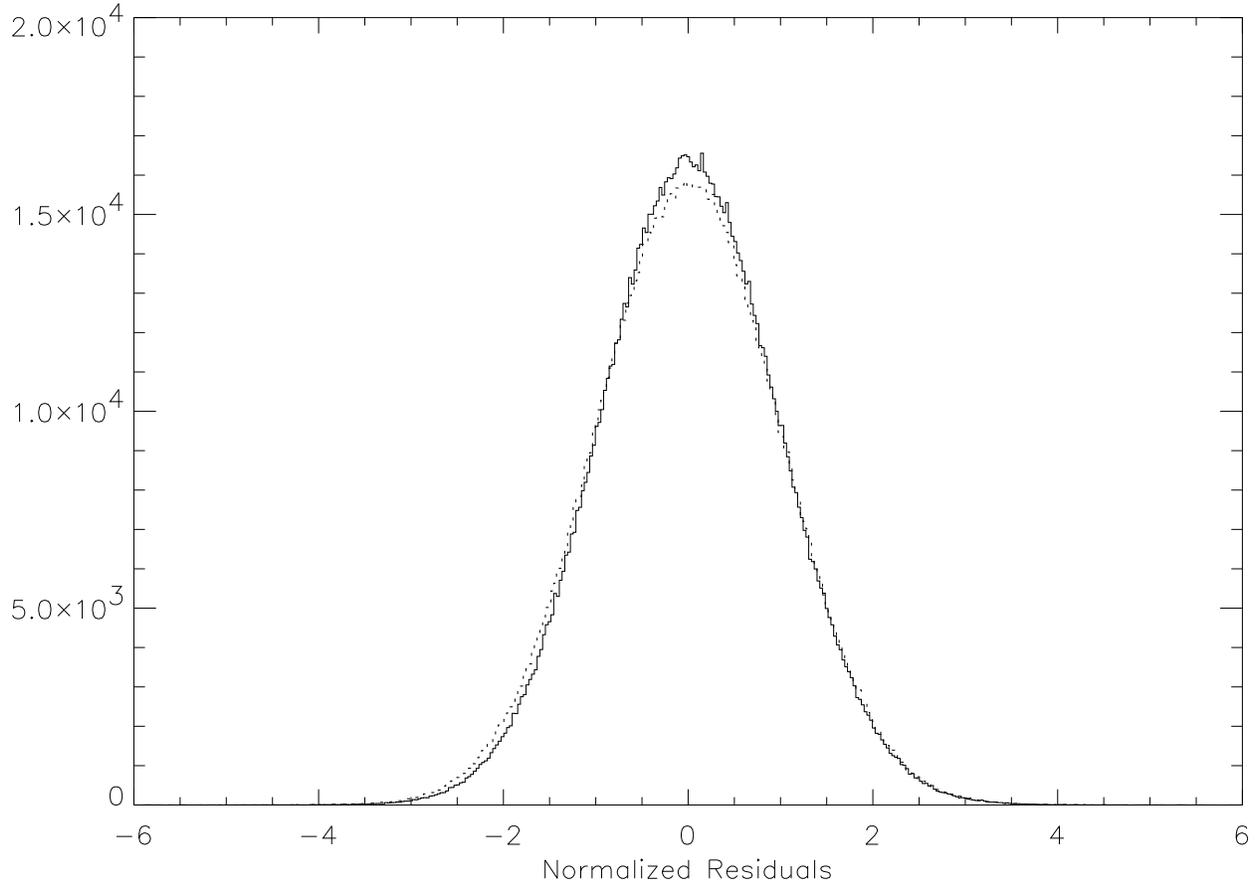}
\caption{The histogram of residuals from a single subtracted image
normalized by the sum of the Poisson errors from the reference image
and data image. Overplotted in a dashed line is the idealized result,
a Gaussian of variance 1 and mean 0. The variance of the subtracted
image is shown to be approximately photon noise.}
\end{figure}

\begin{figure}
\plotone{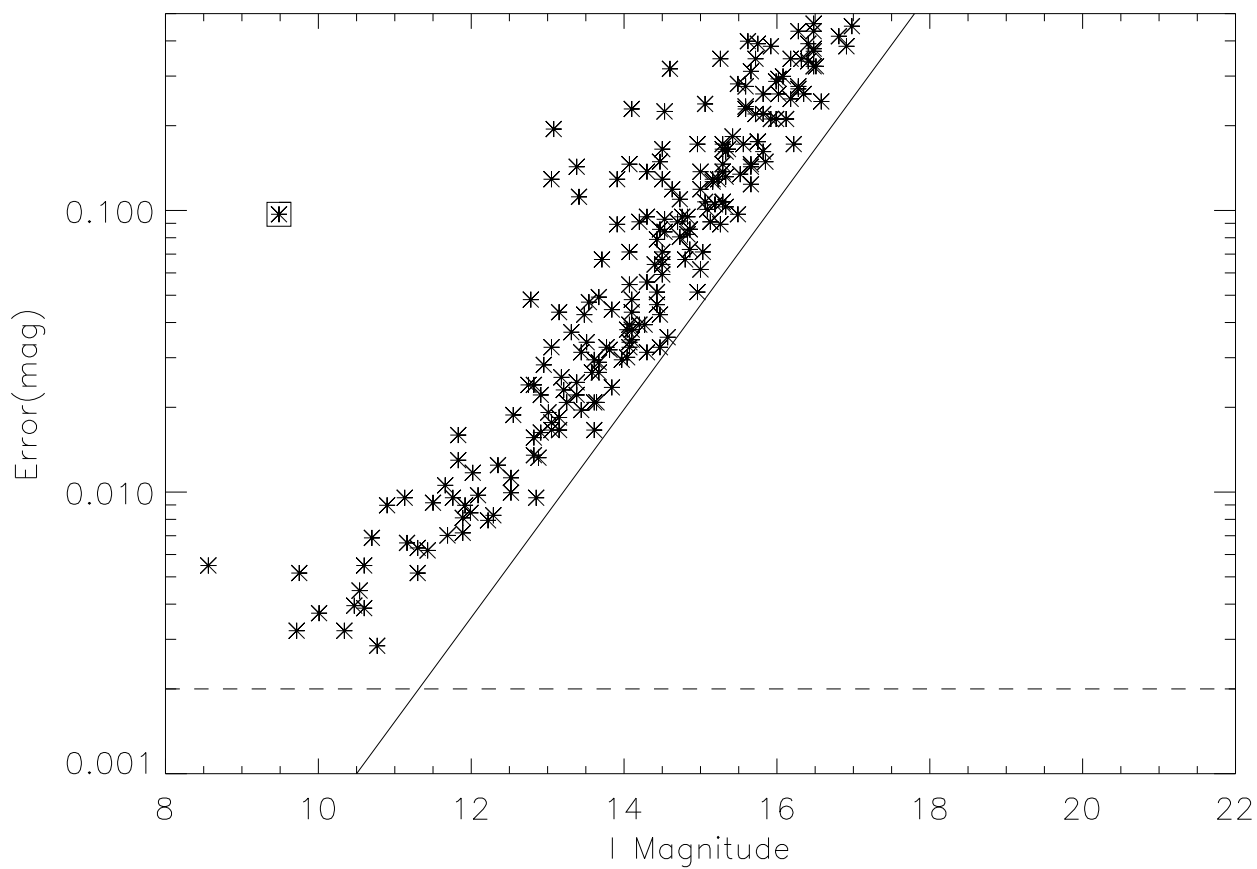}
\caption{Typical errors in magnitudes for stars of a given I band
magnitude. The errors are determined by calculating the ratio of the
standard deviation of stellar flux to average stellar flux for each
star. The overplotted line represents errors that would be expected
from the sum of the the sky background Poisson errors in the reference
and data image. Photometric errors were sky limited for all but
the brightest stars. The dashed line is the theoretical limitation due
to scintillation noise, calculated according to Young (1974).  The
outlying point marked with a box is a bright variable star with light
curve shown in figure 5.}
\end{figure}

\begin{figure}
\plotone{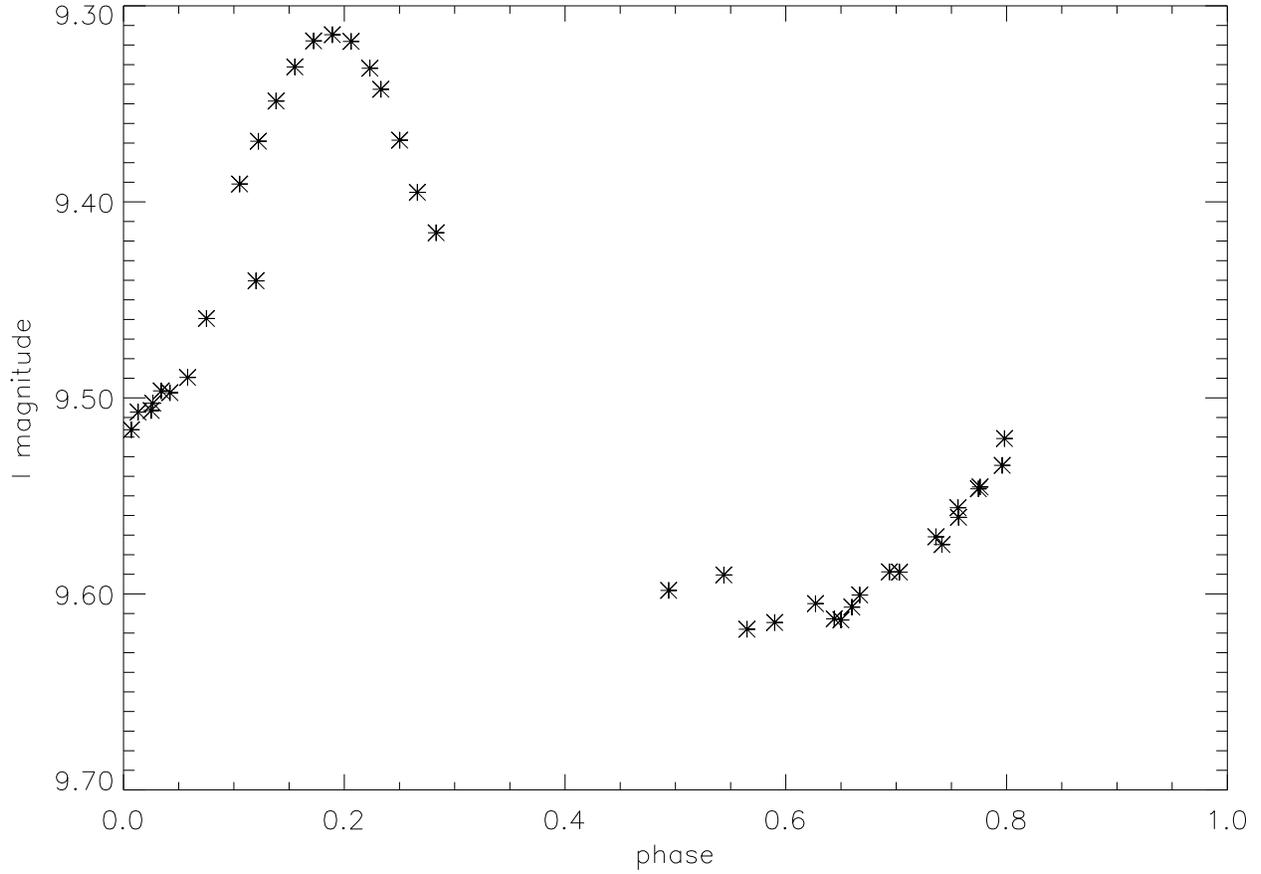}
\caption{The phased light curve of the newly discovered variable star
BD+35 114. The period was found to be 0.81d with an I band amplitude of
0.34 magnitude.  Errors in the magnitudes are approximately equal to
the size of the data points. }
\end{figure}

\end{document}